\documentstyle[aps,prc,epsfig]{revtex}
\begin{document}
\textwidth=17cm
\textheight=22cm
\topskip=-1cm
\footskip=1cm
\footheight=1cm
\oddsidemargin=-1cm %marge de gauche

\title{Fine-Structure of  Choptuik's Mass-Scaling Relation}
\author{Shahar Hod and Tsvi Piran  \\
The Racah Institute for Physics, \\
The Hebrew University, Jerusalem, 91904 Israel \\
}

\maketitle
\par\bigskip

\begin{abstract}

We conjecture (analytically) and demonstrate (numerically) the
existence of a fine-structure above the power-law behavior of the mass
of black-holes that form in gravitational collapse of spherical
massless scalar field \cite{Hod}.  The fine-structure is a periodic function of
the critical-separation $(p-p^*)$. We predict that the period $\varpi$
is {\it universal} and that it depends on the previous universal
parameters, the critical exponent, $\beta$, and the echoing period
$\Delta$ as $\varpi = \Delta /\beta$.

\end{abstract}

\section{Introduction}

The gravitational collapse of a spherically-symmetric massless
scalar-field has two possible end states.  Either the scalar field
dissipates away leaving a flat spacetime or a black-hole forms.
Numerical simulations of this model problem \cite{Choptuik} have
revealed an unexpected critical behavior when the initial conditions
are close to a critical case $p=p^*$ ($p$ is some parameter which
characterizes the strength of the initial scalar field, and $p^*$ is
the threshold value).  More precisely,  Choptuik has found a power-law
dependence of the black-hole mass on critical separation $(p-p^*)$ of
the form: $M_{bh} \propto (p-p^*)^\beta$ for $p>p^*$, and a discrete
echoing with a period, $\Delta$, (a discrete self-similar behavior)
for $p=p^*$.

Subsequently similar critical behavior have been observed for other
collapsing fields: axisymmetric gravitational wave packets
\cite{Abrahams_Evans}, spherically symmetric radiative fluids
\cite{EvansColeman} and complex scalar fields \cite{Hod_Piran}.  In
all these model problems the critical exponent $\beta$ turned out to
be close to the value originally found by  Choptuik 0.37, suggesting a
universal behavior.  However, Maison \cite{Maison} has shown that for
fluid collapse models with an equation of state given by $p = k \rho$
the critical exponent strongly depends on the parameter k.

In this work we conjecture the existance of a small periodic
correction,  $\Psi[(\ln(p-p^*)]$,  to the power-law dependence of the
black-hole mass. $\Psi$ is periodic and its period, $\varpi$, is
universal and it  depends on the previous universal parameters as
$\varpi = \Delta /\beta$. Our analytical argument predicts the
existence of the fine-structure periodic term and its expected period.
The argument is based upon the final stage of a super critical
evolution: from the moment  when the deviation from the exact
self-similar critical evolution becomes larger than some given value
(and the evolution is no longer self- similar) up to the horizon
formation.  We then provide a numerical evidence that verifies the
existence of the conjectured periodic term, the universality of its
period and the relation $\varpi = \Delta /\beta$.

Our numerical formalism is based on the characteristic scheme of
Goldwirth and Piran \cite{Gold_Piran} to which we have added an
expansion near the origin which is essential to achieve the extremely
high accuracy needed for these computations.  The evolution equations,
our algorithm and numerical methods, and the discretization and error
analysis are all described in a previous paper \cite{Hod_Piran}, and
will not be repeated here.

\section{Theoretical Predictions vs. Numerical Results.}

We consider the spherical collapse of a massless scalar field.
Choptuik has shown that for a critical parameter $p^*$ there is a
critical solution which has an infinite discrete self-similar
behavior.  The critical solution by itself does not yield the
black-hole mass scaling relation in which we are interested.  We
perturb, therefore, the critical initial conditions.  This leads to a
dynamical instability - a growing deviation from the critical
evolution toward either sub-critical dissipation or supercritical
black-hole formation.

Let $f(u)$ be a function of $u$, the time coordinate of an observer at
rest at the origin, that characterize the solution along the outgoing
null geodesic that leaves the origin at $u$.  The function, $f$, could
be, for example, the maximal value of $M(r,u)/r$ along this geodesic.
Following Evans and Coleman \cite{EvansColeman} and Maison
\cite{Maison} we describe the run-away of the perturbed solution from
the critical evolution, (described by $f_c$) as a power-law:
\begin{equation} 
f(u)-f_c(u) = \lambda (u^*-u)^{-\alpha} \ ,
\end{equation} 
where the critical solution reaches the zero-mass singularity at
$u=u^*$.  The prefactor $\lambda$ satisfies: $\lambda \propto
(p-p^*)$.

We assume that the range of validity of the perturbation theory is
restricted to some maximal deviation, $\chi$, from the exact critical
evolution, i.e. the evolution is approximately self-similar until
$f-f_c=\chi$.  From here on the evolution is outside the scope of the
perturbation theory - there is sub-critical dissipation of the field
or supercritical black-hole formation.  In either case, the evolution
from this stage onwards looses its self-similar character.  We chose
now $p>p^*$ so that the perturbed initial conditions develop into a
black-hole.  The time $u_\chi(p-p*)$ required in order to reach the
maximal deviation  is given simply by the relation:
\begin{equation}
\lambda (u^*-u_\chi)^{-\alpha} = \chi \ .
\end{equation}
Of course, a larger initial perturbation requires a shorter time to
reach this value.  We define now the logarithmic time, $T \equiv
-\ln[(u^*-u)/u^*]$, in which the critical solution is periodic. The
logarithmic time, $T_\chi$, which corresponds to the loss of
self-similarity, is given by
\begin{equation}
T_{\chi} = -\alpha^{-1} \ln(p-p^*)+b_k \ ,
\label{tci}
\end{equation}
where $b_k$ depends on $\chi$, $u^*$ and $k$. The index $k$ denotes
the family of initial conditions considered.  We conjecture now that
the logarithmic time until the horizon formation, equals to $T_\chi$
plus a periodic term $F[(\ln(p-p^*)]$:
\begin{equation}
T_{bh} = -\alpha^{-1} \ln(p-p^*)+b_k+F[ln(p-p^*)]  \ .
\label{tbh}
\end{equation}
The period, $\varpi$, is universal and it depends on the previous
universal parameters according to:
\begin{equation}
\varpi=\alpha \Delta  \ .
\label{varpi}
\end{equation}

Consider now two different initial-conditions, which lead to n and n+1
echoes respectively (until the deviation from the critical
self-similar evolution reaches $\chi$).  These solutions are related
to each other by an exact scale transformation with a factor
$e^\Delta$.  The final stages of these two supercritical evolutions,
from the stage when the deviation from the exact critical evolution
reaches $\chi$, (and the evolution ceases to be periodic in $T$), up
to the horizon formation, are equal up to a scaling transformation.
The periodic nature of the function F arises from this final stage:
The period of the function F is the amount that should be added to the
quantity $\ln(p-p^*)$, in order to reduce the number of echoes by one.
This will reduce $T_{bh}$ by $\Delta$.  From Eq. \ref{tbh} this amount
to a period $\varpi= \Delta \alpha$ in $F$.

This conjecture is verified by numerical simulations of four families
of initial data (two neutral and two charged).  In all those families
we have found that $-T_{bh}$ as a function of $\ln(p-p^*)$ was well
fit by a straight line with a slope $ 1/\alpha \approx 0.37$.  On top
of this straight line there was a small modulation.  The deviation
from a straight line is shown in Fig. 1. which provides a numerical
evidence for the existence of the periodic term F in Eq. \ref{tbh}.
We see that the function F is indeed periodic, with a universal period
$\varphi = \Delta \alpha \approx 4.6$.

\begin{figure}[ht]
\centerline{\epsfig{file=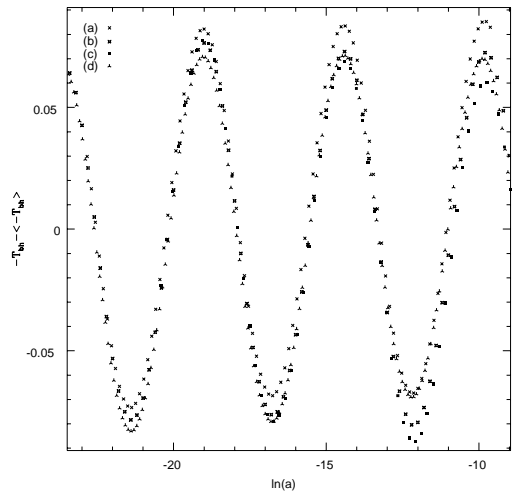,height=200pt}}
\caption{\it Illustration of the conjectured {\it universal
periodic fine-structure } of $-T_{bh}$.  The quantity $[-T_{bh}-
\langle -T_{bh} \rangle] $ is plotted as a function of $\ln(a)$ ,
where $a\equiv (p-p^*)/p^*$ , for the four families.  The curves were
shifted horizontally (but not vertically) in order to overlap the
first oscilation of each family with the first one of family $(a)$.
$\langle T_{bh} \rangle $ is the value of $T_{bh}$ determined from a
straight line approximation, i.e. $\langle T_{bh} \rangle =Const+
\beta \ln(p-p^*)$. The numerical results agree with the {\it predicted
} relation $\varpi= \alpha \Delta \approx 4.6$. }
\end{figure}
We have to relate, now the exponent $\alpha$ to the critical exponent
$\beta $ (which describes the power-law dependence of the black-hole
mass) and then generalize  Choptuik's scaling-relation by proving that
one should also add a periodic term to  Choptuik's mass scaling
relation. We write $T_{bh}$ in the form: 
\begin{equation}
T_{bh} = T_{init}+ n \Delta + F \ ,
\label{tbh2}
\end{equation}
where $T_{init}$ is the initial logarithmic time required for the
system to settles down to a periodic behavior in $T$, and $n$ is the
number of echoes.  We assume that $T_{init}$ is independent of
$(p-p^*)$.  Using Eq. \ref{tbh} we obtain:
\begin{equation} 
n\Delta = -\alpha^{-1} \ln(p-p^*)+d_k \ ,
\label{ndelta}
\end{equation}
where $d_k$ is a family-dependent constant.  We define $M^{(n)}$ as
the mass after $n$ echoes (note that this is not the final black hole
mass). Since $M$ decreases in each echo by a factor $e^{-\Delta}$ we
have, using Eq. \ref{ndelta}:
\begin{equation} 
M^{(n)}=M^{(0)}e^{-n\Delta} = M^{(0)}e^{-d_k}(p-p^*)^{\beta} \ ,
\end{equation}
from which it follows that $\beta = 1/\alpha$.

To obtain, $M_{bh}$,  the final black-hole mass one should multiply
$M^{(n)}$ by a periodic function $G[\ln(p-p^*)]$ which measures the
change of mass, from the stage when the evolution is no longer
periodic in $T$, until the horizon forms.  The function $G$ depends
only on the field configuration at the  moment when the deviation from
the exact self-similar evolution reaches $\chi$  (and the evolution is
no longer self-similar).  Thus, $G$ is expected to have the same value
each time the system completes another echo, i.e.  each time n
increases by unity.  Using Eq. (6) we find that the function
$G[\ln(p-p^*)]$ is expected to have a period of $\varpi=
\Delta/\beta$.  Thus, we obtain 
\begin{equation} 
\ln(M_{bh})=\beta \ln(p-p^*)+c_k+\Psi[\ln(p-p^*)] \ ,
\end{equation}
where  $c_k$  is a family-dependent constant and $\Psi[\ln(p-p^*)]$ 
is a periodic function with a {\it universal} period, $\varpi$.

Fig. 2 depicts  this periodic fine structure for our four families of
solutions mentioned earlier. In all four families we obtain the basic
power law behavior with $\beta \approx 0.37$. Fig. 2 displays the
deviation of $\ln(M_{mb})$ from this straight line as a function of
$\ln(p-p^*)$.  The agreement between the four families shows that the
fine structure is indeed universal with the expected period.  

\begin{figure}[ht]
\centerline{\epsfig{file=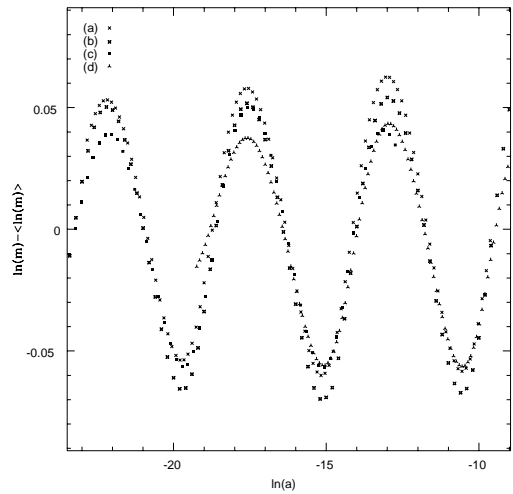,height=200pt}}
\caption{\it Illustration of the conjectured {\it universal
periodic fine-structure } generalization of  Choptuik's mass-scaling
relation.  $\ln(m)-\langle \ln(m) \rangle $ is plotted as a function
of $\ln(a)$ for the four families, where $m \equiv M_{bh}/M_{bh,c}$ is
the normalized black hole mass in units of the initial mass in the
critical solution $M_{bh,c}$.  $\langle \ln(m) \rangle $ is the value
of $\ln(m)$ determined from a straight line approximation. The curves
were shifted horizontally (but not vertically) in order to overlap the
first oscilation of each familiy with the first one of family $(a)$.
The numerical results agree with the {it predected} relation $\varpi=
\Delta /\beta \approx 4.6$.}
\end{figure}
One may worry, of course, whether this fine structure is real or could
it arise from some numerical errors.  In a previous paper
\cite{Hod_Piran} we have established the stability and convergence of
our code with numerous tests. Still becuase of the importance of this
issue we demonstrate  here the physical character of this fine
structure.  Fig. 3 depicts the deviations of $\ln(M_{mb})$ from a
straight line as a function of $\ln(p-p^*)$ for five different
calculations with 100, 200, 400, 800 and 1600 grid points for the same
initial data.  The same features were found on all grids even though
the grid sizes differ by a factor of 16.  The five curves overlap and
all show the same periodic behaviour.  Numerical convergence (the 800
and 1600 curves are nearer than the 100 and 200 curves, for example)
is clearly seen.
\begin{figure}[ht]
\centerline{\epsfig{file=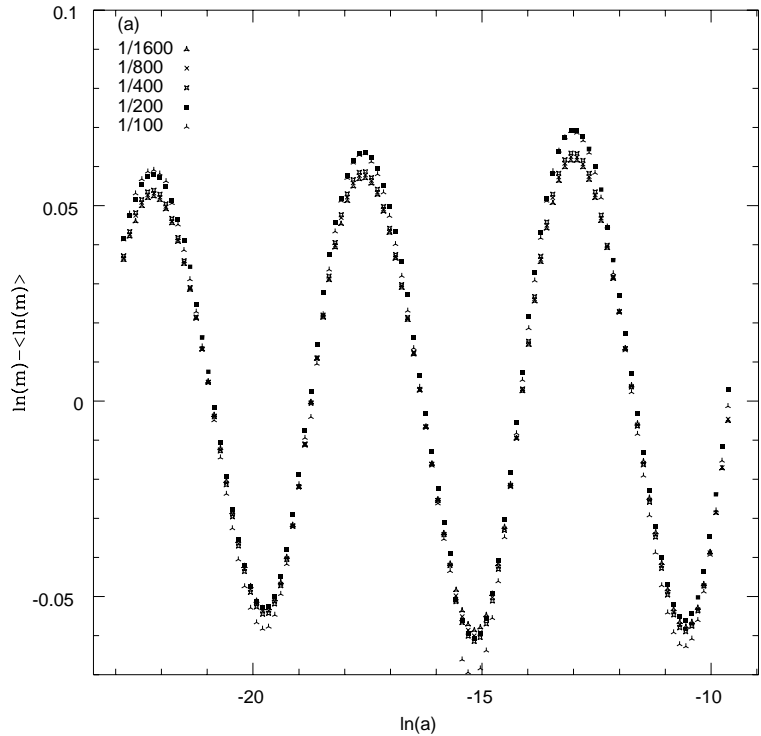,height=300pt}}
\caption{\it $\ln(m)-\langle \ln(m)  \rangle $ is plotted as a
function of $\ln(a)$ for family $(a)$ and for {it five} different
resolution grids with 100,200,400, 800 and 1600 gridpoints. The five
curves overlap and all show the same periodic behavior.}
\end{figure}
\section{Summary and Conclusions.}

We have studied the spherical gravitational collapse of a massless
scalar-field, both for the uncharged case and for the charged
configurations.  Our main interest was the supercritical $(p>p^*)$
feature of  Choptuik's solution, i.e.  the power-law dependence of the
black-hole mass on the  critical separation.   We have shown the
existence of a fine-structure above  this power-law dependence in the
form of a periodic term with a universal period, $\varpi$.  We are not
aware of such fine-structure periodic term in  any other
phase-transitions in statistical mechanics.  Our periodic term with
its period strongly depends on the discrete echoing character of the
critical solution.  This discrete self-similarity has been seen only
in the works of  Choptuik\cite{Choptuik}, Abrahams and Evans
\cite{Abrahams_Evans} (collapse of axisymmetric vacuum gravitational
field) and in our work concerning the gravitational collapse of a
charged (complex) scalar field \cite{Hod_Piran}.  Abrahams and Evans
\cite{Abrahams_Evans} have found an  echoing period of $\Delta \approx
0.3$.  Thus, using our analytical argument, we conjecture that a
careful analysis will reveal a periodic fine-structure (to the
power-law behavior), with a period of $\varpi \approx 0.8$, in the
model problem of the collapse of axisymmetric gravitational wave
packets.

\acknowledgments
We thank Amos Ori for helpful discussions.  This research was
supported by a grant from the US-Israel BSF and a grant from the
Israeli Ministry of Science.

\end{document}